\documentclass{andromedaone}       

\journal{NDM}
\vol{2020}
\jyear{Egypt}
\pages{Hurgada} 

\received{xx January 2018}
\published{xx March 2018}

\newcommand{\ie}{{\it i.e.~}}
\newcommand{\etal}{{\it et al.\,}}
\newcommand{\Qem}{Q_{\rm em}}

\begin{document}

\title{Flavor physics and Jarlskog determinant}

\author{Jihn E. Kim\auno{1,2} }
\address{$^1$Department of Physics, Kyung Hee University, 26 Gyngheedaero, Seoul 02447, Republic of Korea}
\address{$^2$Department of Physics and Astronomy, Seoul National University, 1 Gwanakro, Gwanak-Gu, Seoul 08826, Republic of Korea} 

\begin{abstract}
The flavor problem is reviewed  starting with the chiral symmetry, and  the $A_4$ symmetry derivation  and its realization in GUTs are presented.
\end{abstract}

\maketitle


\section{Introduction}

\bigskip
\noindent
The energy pie of the universe consists of  roughly 68\,\% dark energy (DE), 27\,\% dark matter (DM), and 5\,\% atoms and the rest. Quintessential axions may be some candidates of DE \cite{Carroll98,KimNamIJMPA}, very light axions for DM \cite{PWW83}, and the rest are mostly atoms. In this talk, I will concentrate on atoms which is visible around us. The main reason, that multi-GeV scale atoms have survived the violent condensation process at the Planck and the grand unification (GUT) scales, $10^{16}-10^{18\,}$GeV, is the chirality of the Standard Model (SM) fermions. In the SM, the chirality is introduced by Weinberg \cite{Weinberg67} and formulated in GUTs by Georgi \cite{Georgi79}. Chirality ensures small scales.

\bigskip
\noindent
With the gauge symmetry as the only symmetry at low energy, chiral fields are the only light fields. If more discrete or global symmetries are imposed beyond the gauge symmetries, one can obtain almost infinite number of possibilities for light fields. In view of the fact that TeV scale particles have not been observed so far at the Large Hadron Collider (LHC), the extra global or discrete symmetries are not needed at the moment. It is very rare to have simple chiral models. So, I show one interesting chiral model based on SU(2)$'\times$U(1)$'_Q$ in the hidden sector, which was found recently \cite{Kim17}, 
\begin{eqnarray}
Q'&=&\frac12:~ \ell_i=\left(\begin{array}{c}E_i\\[0.3em] 
N_i \end{array}\right)_{\frac{+1}{2}},~~\begin{array}{c}E_{i,-1}^c\\[0.4em] 
N_{i,0}^c \end{array}~~~~~(i=1,2,3),\\[0.7em]
Q'&=&-\frac32:~ {\cal L}=\left(\begin{array}{c}{\cal E}\\[0.3em] 
{\cal F}\end{array}\right)_{\frac{-3}{2}},~~\begin{array}{c}{\cal E}_{,+1}^c\\[0.4em] 
{\cal F}_{,+2}^c \end{array}
\end{eqnarray}
So, there is a good chance that these particles will be discovered at low energy, mainly through  the kinetic mixing with the SM photon and $Z^0$ gauge boson \cite{Kim17}.

\begin{figure}[b!]
\hskip 3cm
\includegraphics[height=4cm]{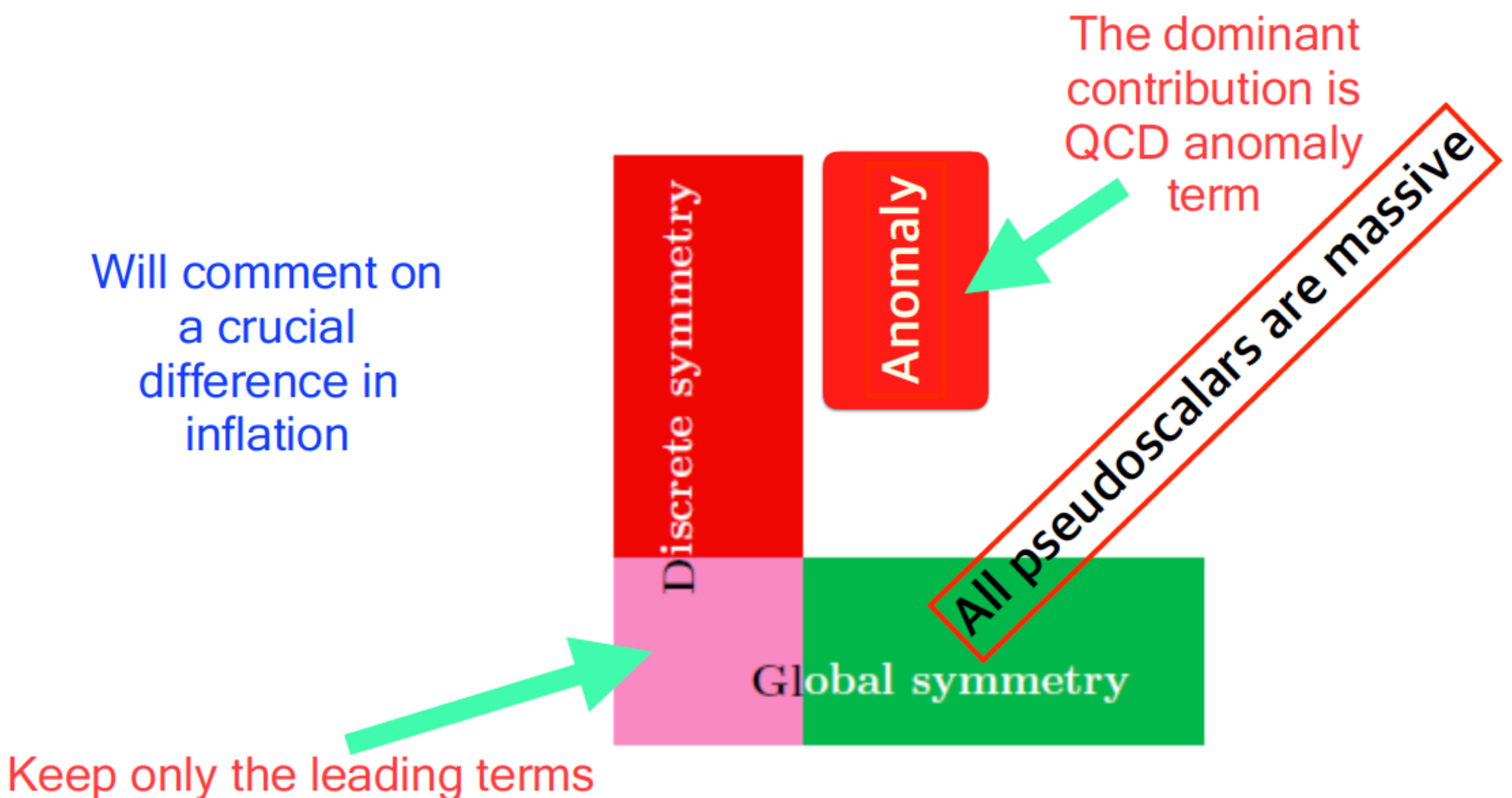}
\caption{An approximate global symmetry from few terms allowed by a discrete symmetry.}
\label{f:DiscGlobal}
\end{figure}

\bigskip
\noindent
I have worked on models related to chiral symmetries for a long time. One is the very light axion which arise by breaking a chiral symmetry at an intermediate scale \cite{Kim79}, much above the electroweak (EW) scale.  From a top-down approach, e.g. from superstring compactification down to 4 space-time dimensions (4D), there can be discrete symmetries even though global symmetries are forbidden.
The terms allowed by such a discrete symmetry is shown as the far left vertical column in Fig. \ref{f:DiscGlobal}. If we keep only a few dominant terms, \ie the terms allowed in the lavender square,  there can be a global symmetry. The terms symbolized by the red break that global symmetry.
  There are two sources for breaking that global symmetry. One is the terms ($\Delta V$) neglected in the potential $V$, which is marked as the red in the far left column. The other is the anomaly which is marked as the other red. For the strong CP solution, this anomaly called the Peccei--Quinn anomaly is the only violation of the global symmetry without the violation from $\Delta V$. For the anomaly violation, it is known that the minimum of the potential of the axion is at $\bar{\theta}=a/f_a=0$ \cite{VafaWitten84}.

\begin{figure}[h!]
\hskip 2cm
\includegraphics[height=6cm]{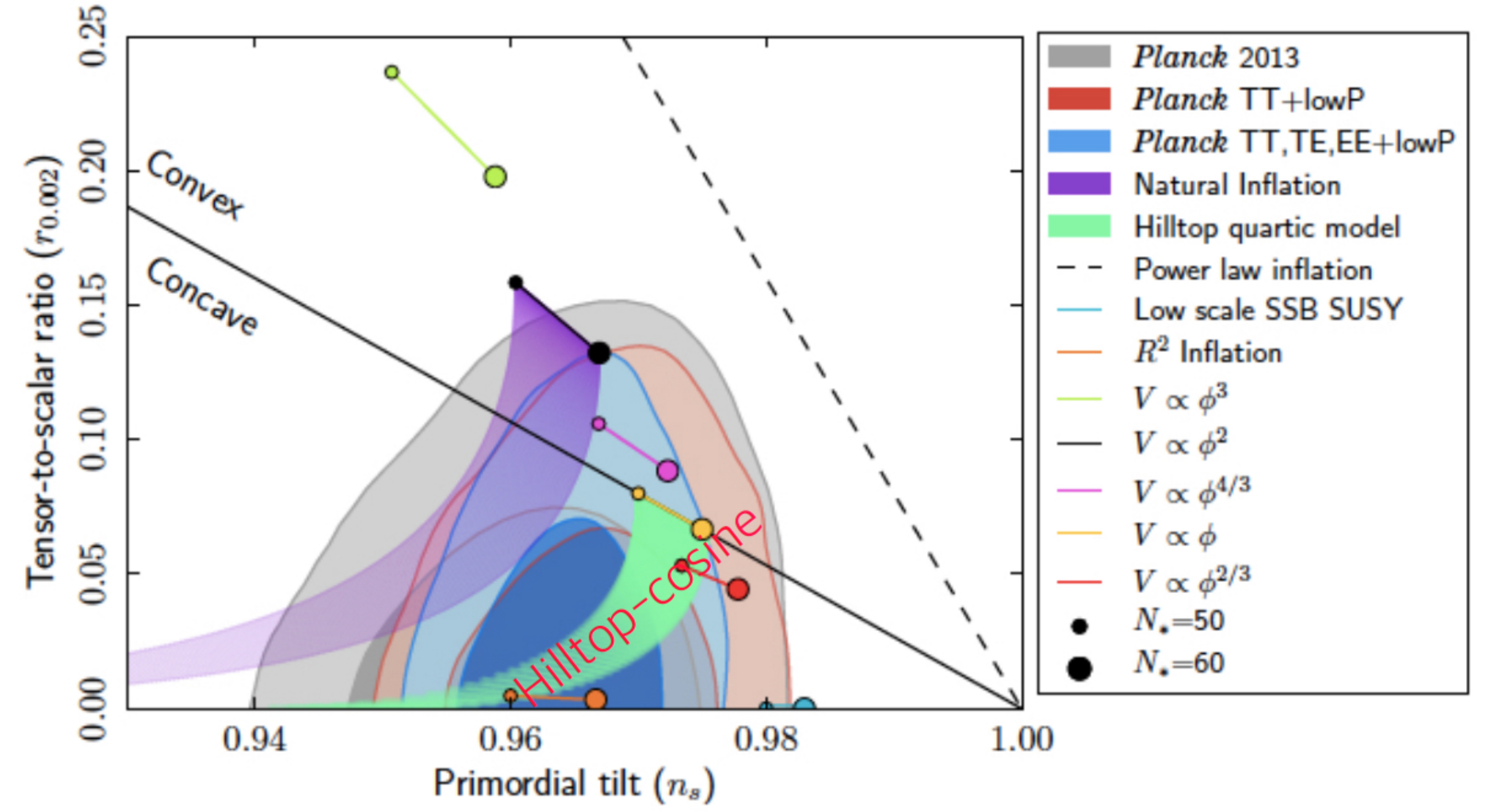}
\caption{The tensor to scalar ratio versus $n_s$. The hilltop-cosine model is marked as the bright green.}
\label{f:Tns}
\end{figure}

\bigskip
\noindent
Natural inflation \cite{Freese92} uses this idea of breaking global symetries by gauge anomalies. The natural inflation prediction on the tensor to scalar ratio and $n_s$ are shown as the violet band in Fig. \ref{f:Tns}. However, as commented in Fig.  \ref{f:DiscGlobal}, the global symmetry can be broken also by $\Delta V$ in which case the origin can be a maximum by an appropriate choice of parameters in $\Delta V$. In this case, ``natural inflation'' is a kind of hilltop inflation, which is marked as green {\it Hilltop-cosine} in Fig. \ref{f:Tns}. In any case, old natural inflationists have not suggested appropriate nonabellian gauge groups for making the origin as the minimum. In string compactification, most probably the natural inflation uses $\Delta V$ for breaking the global symmetry simply because string compactification may not produce the needed nonablelian gauge groups.

\section{Flavor problem}
 
\bigskip
\noindent
Let us now focus on our main problem: the flavor problem.
One family of the SM is
\begin{eqnarray}
&&Q_L^\alpha,~~ (u^\alpha)^c_L,~~  (d^\alpha)^c_L,~~\ell_L, ~e_L^c,\label{eq:SM1family}
\end{eqnarray}
where
\begin{eqnarray}
&& Q_L^\alpha=\left(\begin{array}{c}u^\alpha\\[0.3em] 
d^\alpha\end{array}\right)_L,~~~~\ell_L=\left(\begin{array}{c}\nu_e \\[0.3em] 
e \end{array}\right)_L
\end{eqnarray}
where $\alpha=1,2,3$ is the color  index. The family structure, fifteen chiral fields of Eq. (\ref{eq:SM1family}), repeats two more times, in total three families. The family problem started with the 2nd family when I. I. Rabi quipped ``Who ordered that?'' in 1937 in response to the news that the 1936-discovered muon was not a hadron but a new and entirely unexpected type of lepton. Also, Cabibbo's mixing angle started with the $K$ meson decay, and the atmospheric neutrino oscillation was the oscillation of muon-neutrinos. But the essence of the family problem starts with the 3rd family whose members are the heaviest. 

 \bigskip
\noindent
 It is our freedom to choose the bases of quark and leptons. Let us choose the bases where $\Qem=-\frac13$ quark and $\Qem=-1$ lepton masses are already diagonalized. These masses arise from the renormalizable couplings with a $Y=-\frac12$ Higgs doublet. Then, the  $\Qem=+\frac23$ quark masses span the $3\times 3$ quark mass matrix in the flavor space. For neutrino masses, it occur at the nonrenormalizable level, which is not of our concern here. The weak eigenstates with superscipt $^{(0)}$ are related to the mass eigenstates by a unitary transformations,
 \begin{eqnarray}
  \left(\begin{array}{c}u^{(0)}\\[0.3em] 
c^{(0)}\\[0.3em] t^{(0)}\end{array}\right)_L= A_L^{(u)\dagger}\left(\begin{array}{c}u\\[0.3em] 
c\\[0.3em] t\end{array}\right)_L;~~ \left(\begin{array}{c}\nu_{e}^{(0)}\\[0.3em] 
\nu_{\mu}^{(0)}\\[0.3em] \nu_{\tau}^{(0)}\end{array}\right)_L= B_L^{(\nu)\dagger} \left(\begin{array}{c}\nu_{e}\\[0.3em] 
\nu_{\mu}\\[0.3em] \nu_{\tau}\end{array}\right)_L,
\end{eqnarray}
and similarly for the R-hand components for quarks with $A_R^\dagger$. In our case, the CKM matrix is $A_L^{(u)}A_L^{(d)\dagger}=A_L^{(u)}$. The 9 parameters of $A_L^{(u)}$ describe three quark masses, two unobservable phase degrees of quarks (one cannot be used for phas redefinition because of the baryon number conservation), three real angles and one phase. The PMNS matrix is $B_L^{(\nu)}B_L^{(e)\dagger}=B_L^{(\nu)}$. The 9 parameters of $B_L^{(\nu)}$ describe three neutrino masses, two observable Majorana phases,  three real angles and one Dirac phase. In the PMNS case, one cannot rotate away some phases because Majorana fields are basically real.

\bigskip
\noindent
In Ref. \cite{FK20}, we argued that CKM matrix is close to the identity due to the hierarchy of quark masses. There is no such huge hierarchy for neutrino masses.  But the PMNS matrix needs not be close to the identity. If some non-diagonal elements of the PMNS matrix are the same, usually discrete symmetry is assumed. If 4 elements in the PMNS matrix are the same, there must be a permutation symmetry with 4 objects, \ie $S_4$. Furthermore, suppose that one element of the PMNS matrix is zero as shown below
\begin{eqnarray}
  \left(\begin{array}{ccc} \times &  \times &  \times \\[0.3em] 
 \times &  \times &  \times \\[0.3em] 
 0 & \times &  \times \end{array}\right) \to  
 \left(\begin{array}{ccc} 
 \times &  \times &  \times \\[0.3em] 
 p &  q &  r \\[0.3em] 
 {\bf S}& \times &  \times \end{array}\right) .\label{PMNS0}
\end{eqnarray}
To study the matrix with one entry being 0, let us consider the RHS form with $p,q,r$ and ${\bf S}$ are the same. Permutations of  $p,q,r$ and ${\bf S}$ constitute the 24 element group $S_4$, which are shown below,
\begin{eqnarray}
{\rm Cyclic}:\quad&& pqr {\bf S},~qrp {\bf S},~rpq {\bf S},\nonumber\\
&&pq {\bf S} r,~qr  {\bf S} p,~rp {\bf S} q,\nonumber\\
&&p {\bf S} qr,~q  {\bf S} rp,~r  {\bf S} pq,\label{eq:tvplus}\\
 &&{\bf S} pqr,~  {\bf S} qrp,~ {\bf S} rpq, \nonumber
\end{eqnarray}
 and
\begin{eqnarray}
{\rm Anticyclic}:&&  qpr {\bf S},~prq {\bf S},~rqp {\bf S},\nonumber\\
&&qp{\bf S} r,~pr {\bf S} q,~rq {\bf S} p,\nonumber\\
&&q  {\bf S}pr,~p  {\bf S}rq,~r   {\bf S}qp,\label{eq:tvminus}\\
&& {\bf S}qpr,~  {\bf S}prq,~ {\bf S}rqp.\nonumber
\end{eqnarray}
The 24 elements of $p,q,r$, and ${\bf S}$ are assumed to be identical because of the $S_4$ symmetry. These 24 elements can be identical if we allow ${\bf S}=+p$ and also  ${\bf S}=-p$ but consider only the cyclic one , Eq. (\ref{eq:tvplus}), \ie   ${\bf S}=-p$ with Eq. (\ref{eq:tvplus}) reproduces  Eq. (\ref{eq:tvminus}). Now consider  Eq. (\ref{eq:tvplus}) and set ${\bf S}=0$ by hand. Then, the 24 elements collapses to only 12 elements. The case  ${\bf S}=0$ reproduces the LHS of Eq. (\ref{PMNS0}), and we proved that the tetrahedral group $A_4$ with 12 elements is the symmetry for the trimaximal mixing with one entry being 0.  Note that we have not proved `tri-bimaximality' but only `trimaximality' at this stage. To prove `tri-bimaximality', one needs one more criterion.

\section{CKM and PMNS matrices:~J confronts data}
  
  \bigskip
  \noindent
Most flavor physics start from the CC weak interactions since the discovery of  $\beta$ decay. Usually, the CKM matrix is defined by the charge raising and the PMNS matrix by the charge lowering  CC operators \cite{PDG18}.
For a unified description of the CKM and PMNS matrices, it is better to use one kind of CCs. Here, we use the charge raising CCs, and hence our PMNS matrix is the hermitian conjugate of the matrix given in the PDG book \cite{PDG18}. So, to compare the CP phase in the PMNS matrix of \cite{PDG18} with the phase in the CKM matrix in a unified description, one should take another minus sign for the PMNS phase. The Jarlskog determinant $J$ is read from the CKM matrix itself as discussed in Ref. \cite{KimSeo12}, after making Det.$V^{\rm CKM}$=real,
\begin{equation}
J=\Big|{\rm Im}V^{\rm CKM}_{31}V^{\rm CKM}_{22}V^{\rm CKM}_{13}\Big|
\end{equation}
 So, an approximate form for the CKM matrix is not accurate enough to discuss $J$ of O($10^{-5}$).
The Kim-Seo (KS) parametrization for the CKM matrix is
\begin{eqnarray}
V^{\rm CKM,KS}= \left(\begin{array}{ccc} c_1,&s_1c_3,& s_1s_3  \\ [0.2em]
-c_2s_1, & c_1c_2c_3+ s_2s_3 e^{-i\delta},&c_1c_2s_3-s_2c_3e^{-i\delta}  
 \\[0.2em]
 -s_1s_2e^{+i\delta},&  -c_2s_3 +c_1s_2c_3 e^{+i\delta}, & c_2c_3 +c_1s_2s_3 e^{+i\delta}
\end{array}\right),\label{eq:UKS}
\end{eqnarray}
where $c_i$ and $s_i$ are cosine ans sine of the real angle $\theta_i$ and $e^{i\delta}$ is the weak CP phase. The PDG book presents the data for each element in terms of absolute values. These central values are not the values satisfying the unitarity condition. The absolute values in (\ref{eq:UKS}) which do not depend on the phase $\delta$ are for the elements (11), (12), (13), (21), and (31) elements. [Similarly, 5 elements for the PDG parametrization do not depend on the phase.] Since we want to express the experimentally determined unitary matrix with the phase $\delta$ undetermined, we use these 5 elements to determine three real angles.
With this philosophy, we determined a data-guided unitary matrix, \ie determine
angles  $\theta_i$ which produce a unitary
matrix for central values. We also present errors of these angles around the central values.
\begin{equation}
V^{\rm CKM; KS}=\left(
\begin{array}{ccc}
 0.975188\pm 0.0013167 \,, & 0.221345\pm 0.00579925\,, & 0.003888\pm 0.000307\\[1.5em]
-0.221226\pm 0.00579616\,,  & \begin{array}{l}0.974365\pm 0.00131633 \\[0.2em] 
 -i(0.00065\pm 0.000003)\sin\delta\\[0.2em]
  +(0.00065\pm 0.000003)\cos\delta \end{array}, &  \begin{array}{l} +0.01712  \pm 0.000002\\[0.2em] 
  +i(0.03712\pm 0.001473)\sin\delta\\[0.2em]
  -(0.03712\pm 0.001473)\cos\delta \end{array}   \\[2.2em]
  \begin{array}{c}   -i(0.00822\pm 0.000482)\sin\delta \\
  -(0.00822\pm 0.000482)\cos\delta
   \end{array} , & \begin{array}{l} -0.017551\pm 0.000002\\[0.2em] +i(0.03620\pm 0.001379)\sin\delta\\[0.2em] +(0.03620\pm 0.001379)\cos\delta \end{array}, & \begin{array}{l} +0.999156\pm 0.00007\\[0.2em] + i(0.00064\pm 0.00005)\sin\delta \\[0.2em] + (0.00064\pm 0.00005)\cos\delta \end{array} 
\end{array}\right) \label{eq:aCKMdata}
\end{equation}
which gives the following real angles in the KS parametrization
\begin{eqnarray}
\theta_1=12.79^{\rm o}, ~\theta_2=2.128^{\rm o}, ~\theta_3=1.006^{\rm o},  
\end{eqnarray}
and we obtain $J=(3.114\pm 0.0325)\times 10^{-5}|\sin\delta|$.

\bigskip
\noindent
It is important to find a progenitor quark mass matrix with the following ansatz in units of $m_t$ \cite{FK20},
\begin{equation}
M=\left(
\begin{array}{ccc}
 q\,\varepsilon^2 , & f\,\varepsilon^{3/2}, & c\,\varepsilon\\[.5em]
 g\,\varepsilon^{3/2}\,,  & p\,\varepsilon , &   a\,\varepsilon^{1/2} \\[0.5em]
  d\,\varepsilon , & b\,\varepsilon^{1/2}, & 1
  \end{array}\right) .
\end{equation}
If data guided mass matrix takes the form, 
\begin{equation}
M=\left(
\begin{array}{ccc}
O(0.00005) , &\ll O(0.0006), & O(0.007)\\[.5em]
O(0.0006),  &O(0.007) , &\ll O(0.084) \\[0.5em]
\ll O(0.007) , &O(0.084) & 1
  \end{array}\right),
\end{equation}
then one can assume some texture zero entries, for example,
\begin{equation}
M=\left(
\begin{array}{ccc}
O(0.00005) , &0, & O(0.007)\\[.5em]
O(0.0006),  &O(0.007) , &0 \\[0.5em]
0 , &O(0.084) & 1
  \end{array}\right),
\end{equation}
In this sense, it is very useful to know the progenitor quark mass matrix as accurately as possible, which will be shown later \cite{Kim20}.

\bigskip
\noindent
The experimental data for the PMNS matrix is usually presented in terms of angles, which guarantees the unitarity of the PMNS matrix. At present, it is not known whether neutrino masses are normal hierarchy (NH) or inverted hierarchy (IH). Also, the data are analyzed with or without including SK atmospheric data. For example,  the NH without the SK atmospheric data gives the Jarlskog determinant  $J=(2.96\pm0.29)\times 10^{-2}|\sin\alpha_{\rm KS}|$.
  
\section{A model in anti-SU(5)}
String compactification does not need a GUT group for unification of coupling constants, which had led to proliferation of standard-like models in string compactification \cite{KimRp19}. One merit of the Georgi-Glashow SU(5) is the $b-\tau$ mass unification at the GUT scale \cite{BEGN77}. If one really wants GUTs from string compactification, the GUT group is better to be a Pati-Salam-type or an anti-SU(5). These anti-SU(5) GUT groups do not necessarily need adjoint representation(s) for the Higgs mechanism to break the GUT group SU(5)$\times$U(1)$_X$. The Georgi--Glashow SU(5) and SO(10) GUTs require adjoint representations for breaking the GUTs. However, for anti-SU(5) \cite{DKN84} (or flipped-SU(5) \cite{Barr82}), the antisymmetric tensor Higgs fields can break the rank-5 SU(5)$\times$U(1)$_X$ down to the rank-4 SM because ${\bf 10}_{-1}$ and $\overline{\bf 10}_{+1}$ contain electromagnetically neutral components and reduces the rank by one unit.   The anti-SU(5) was discussed at the field theory level in \cite{Barr82, DKN84} and in string compactification in \cite{Ellis89,Kim07}. 

  \bigskip
  \noindent
  In particular, we noted in \cite{Huh09} that the GUT group factors are SU(5)$_{\rm visible}\times $SU(5)$'_{\rm hidden}\times$U(1)'s. It is possible to consider that SU(5)$'_{\rm hidden}$ is the confining force in the hidden sector. Indeed, this choice seems plausible for dynamical SUSY breaking as proved in \cite{KimPLB19}. The representation under  SU(5)$'_{\rm hidden}\times$SU(2)$_{\rm global}$ is chiral,
  \begin{equation}
 ({\bf 10}',{\bf 1})\oplus ({\bf 5}',{\bf 1}) \oplus (\overline{\bf 5}',{\bf 2})\oplus ({\bf 1}',{\bf 2}),
  \end{equation}
  and one can write a superpotential term 
\begin{equation}
 W\sim ({\bf 10}')^{\alpha\beta}({\bf 10}')^{\gamma\delta} ({\bf 5}')^{\epsilon} \varepsilon_{\alpha\beta\gamma\delta\epsilon}.
  \end{equation}
Below the confinement scale of SU(5)$'$, it was shown that the composite field 
\begin{equation}
\Phi\equiv ({\bf 10}')^{\alpha\beta}({\bf 10}')^{\gamma\delta} ({\bf 5}')^{\epsilon} \varepsilon_{\alpha\beta\gamma\delta\epsilon} 
  \end{equation}
does not satisfy the SUSY conditions and SUSY is broken dynamically. The confining scale is at the intermediate scale and the dynamical SUSY breaking scale and the ``invisible'' axion scale are at the same region as anticipated long time ago \cite{Kim84}. In the anti-SU(5) model, it was possible to study the flavor problem.

\bigskip
\noindent
The flavor problem in GUTs is better to define CCs with one kind of  charge raising or lowering operators to unify the CCs of the quark and lepton sectors. We chose the charge raising operators for defining the CKM and PMNS matrices. In the anti-SU(5), we find one kind of $A_4$ representations in the leptonic sector \cite{FK20} such that the $A_4$ multiplication of two lepton representations satisfy
\begin{equation}
{\bf 3}\otimes {\bf 3}= 2\cdot {\bf 3}\oplus {\bf 1}\oplus {\bf 1'}\oplus {\bf 1}''.
\end{equation}
The reason for this product is chosen is that the mass matrix must belong to   ${\bf 1}\oplus {\bf 1'}\oplus {\bf 1}''$ such that three parameters can be arbitrary, \ie
 three independent electron-type  lepton masses from
  \begin{equation}
  \left( \begin{array}{ccc} M_{11}& M_{12}& M_{13}\\
 M_{21}& M_{22}& M_{23}\\
 M_{31}& M_{32}& M_{33}\\
  \end{array} \right)
   \end{equation} 
where all $M_{ij}$ can be different.
 
\section{Conclusion}
  \noindent
  I reviewed the flavor problem with an emphasis on the $A_4$ symmetry and its implementation in GUTs.
  
\section*{Acknowledgements}
I thank Shaaban Khalil for inviting me to this interesting conference. This work is supported in part  by the National Research Foundation (NRF) grant  NRF-2018R1A2A3074631.

\bibliographystyle{unsrt}

\end{document}